\documentstyle[aps,pra,subfigure,epsfig,amsmath,amssymb]{revtex}

\begin{document}

\draft

\title{Hydrodynamic Excitations in a Spin-Polarized Fermi Gas under Harmonic
Confinement in One Dimension}

\author{ A. Minguzzi, P. Vignolo, M.~L. Chiofalo, and
M.~P. Tosi} 

\address{Istituto Nazionale per la  Fisica della Materia and
Classe di Scienze, Scuola Normale Superiore, \\ Piazza dei Cavalieri 7,
56126 Pisa, Italy \\}

\maketitle                
\begin{abstract}
We consider a time-dependent non-linear Schr\"odinger equation in one
dimension (1D) with a fifth-order interaction term and external
harmonic confinement, as a model for both ({\it i}) a Bose gas with
hard-core contact interactions in local-density approximation, and ({\it ii}) a
spin-polarized Fermi gas in the collisional regime. We evaluate
analytically in the Thomas-Fermi limit the density fluctuation
profiles and the collective excitation frequencies, and compare the
results for the low-lying modes with those obtained from the numerical
solution of the Schr\"odinger equation. We find that the excitation
frequencies are multiples of the harmonic-trap frequency even in the
strong-coupling Thomas-Fermi regime. This result shows that the
hydrodynamic and the collisionless collective spectra coincide in the
harmonically confined 1D Fermi gas, as they do for sound waves in its
homogeneous analogue. It also shows that in this case the
local-density 
theory reproduces the exact collective spectrum of the hard-core Bose
gas under harmonic confinement.
\end{abstract}

\section{Introduction}
The study of atomic gases in condensed quantum states inside magnetic
or optical traps has received an enormous impulse from the achievement
of Bose-Einstein condensation in vapors of bosonic alkali atoms and of
atomic hydrogen \cite{first_exp}. Similar techniques of trapping and
cooling are being used to bring gases of fermionic alkali atoms into
the quantum degeneracy regime \cite{ferm_exp}. The observation of
small-amplitude shape-deformation modes in Bose-Einstein condensates
inside three-dimensional (3D) anisotropic traps at extremely low
temperatures \cite{coll_exp} has proved to be an important method of
diagnostics and has provided a crucial test of the mean-field theory
based on the time-dependent Gross-Pitaevskii equation
\cite{stringari_96}. No such experiments have as yet been observed in
Fermi gases, but predictions are available on their dynamical spectra
in both the collisional and the collisionless regime
\cite{fermi3d}. For both bosons and fermions in 3D the strong-coupling
Thomas-Fermi limit is characterized by a different spectrum from the
weak-coupling limit. 

Attention has also been moving towards atomic gases in restricted
geometries, where new frontiers are met in the realization of atom
lasers \cite{atomlaser,anna_paris} and of thin atom waveguides
\cite{olshanii}. The rich phase diagram of a quasi-1D Bose gas may
become accessible to observation through tuning of the interactions
\cite{slhyap}. Essential modifications to the Gross-Pitaevskii
equation are needed in D$\leq 2$, since the T-matrix vanishes already
in D=2. The dilute gas limit can still imply strong coupling in 1D,
since collisions are unavoidable in this reduced dimensionality. 

At low temperature and density the dynamics of a Bose gas with
repulsive interactions inside a very thin waveguide approaches that of
a 1D fluid of impenetrable point-like bosons \cite{olshanii}. As
demonstrated by Girardeau \cite{girardeau}, this so-called Tonks gas  
\cite{tonks} has the same spatial profiles as those of a
spin-polarized Fermi gas, since the exact many-boson wavefunction is
related to that of non-interacting spinless fermions by 
$\Psi_B(x_1,...,x_N)=|\Psi_F(x_1,...,x_N)|$. More generally, the
Fermi-Bose mapping theorem implies that the dynamic structure factor
of the Tonks gas is the same as that of the corresponding ideal Fermi
gas, although their momentum distributions are very different. In
particular, in the same work Girardeau \cite{girardeau} also showed
that long-wavelength sound waves propagate in the homogeneous Tonks
gas with velocity $v=\pi\hbar n/m$ equal to the Fermi velocity $v_F$
in the homogeneous ideal Fermi gas at the same density $n$ and atom
mass $m$. This is, in fact, the speed of propagation for both zero
(collisionless) sound and hydrodynamic sound in the ideal 1D Fermi
gas. 

The dynamic structure factor of the spin-polarized 1D Fermi gas under
harmonic confinement has been evaluated exactly in the collisionless
regime and related to that of the homogeneous gas by means of a
local-density formula \cite{VMT}. The collective excitation
frequencies in this regime are integer multiples of the harmonic trap
frequency and this spectrum is, therefore, also that of the
corresponding Tonks gas. In the present work we evaluate the dynamics
of the same Fermi system in the collisional regime described by
linearized hydrodynamic equations. We shall show how these equations
for the Fermi gas can be derived from the non-linear Schr\"odinger
equation proposed by Kolomeisky {\it et al.} \cite{kolomeiski} for the
dynamics of the mesoscopic wave function of the corresponding Tonks
gas and how the equality of the spectra of the two fluids remains
valid within the dynamical regime of present interest.

After a presentation of the model in Section II, we evaluate its
dynamical density fluctuations and collective frequency spectrum in
the following Sections. We give in Section III an analytic solution of
the equations of motion for the density fluctuations, using a
Thomas-Fermi approximation (TFA) extended, however, to consider
solutions external to the classical radius. In Section IV we
treat the same problem by means of a numerical simulation mimicking an
experimentally realizable method for measuring the frequencies of the
gas. That is, we excite a collective mode of the atomic cloud by
applying an external time-dependent potential of chosen symmetry and
probe it by observing the evolution in time of the density profile
once the perturbation is turned off. Finally, Section V gives a brief
summary of our main results and some concluding remarks.

\section{The model} 
According to Kolomeisky {\it et al.} \cite{kolomeiski}, a long-wavelength
approach to the hard-core 1D Bose gas with repulsive point-like interactions
in the dilute regime can be based on the non-linear Schr\"odinger
equation
\begin{equation}
i\hbar \partial_t \Phi(x,t)=\left[-\frac{\hbar^2}{2m}
\partial_x^2 +V(x,t)+\frac{\pi^2\hbar^2}{2m}
|\Phi(x,t)|^4  \right] \Phi(x,t)\;.
\label{gp4}
\end{equation}
Here, $V(x,t)=V_{ext}(x)+U_p(x,t)$ is a time-dependent external
potential, 
which includes  the harmonic confinement $V_{ext}(x)=m\omega_{ho}^2 x^2/2$ and
 a periodic perturbation $U_p(x,t)$. The wave function
$\Phi(x,t)$ is normalized to the number $N$ of particles in the trap. 
Equation~(\ref{gp4}) takes account of the correct density dependence of the
ground-state energy density for the 1D Bose gas in the strong-coupling
limit within a
local-density approximation. 

It was shown by Girardeau and Wright
\cite{girardeau_td} that Eq.~(\ref{gp4}) describes
well the dynamics of the Bose gas, although it overestimates its
coherence in interference patterns at small numbers of
particles.
Equation~(\ref{gp4}) can be cast in the form of Landau's hydrodynamic
equations for a superfluid by setting
$\Phi(x,t)=\sqrt{n(x,t)}\exp[i\phi(x,t)]$. This yields  
\begin{equation}
\partial_t n(x,t) =-\partial_x \left[n(x,t) v(x,t)\right]
\label{cont}
\end{equation}
and
\begin{equation}
m\partial_t v(x,t)=-\partial_x \left[\mu_{loc}(x,t)+\frac{1}{2}m
v^2(x,t)\right]\; ,
\label{eqv}
\end{equation}
where the velocity field  is defined as $v(x,t)=(\hbar/m)\partial_x \phi(x,t)$
and the local chemical potential is given by
\begin{equation}
\mu_{loc}(x,t)=-\frac{\hbar^2}{2m\sqrt{n(x,t)}}\partial_x^2
\sqrt{n(x,t)} + V(x,t) + \frac{\pi^2\hbar^2}{2m}
n^2(x,t)\;.
\label{muloc}
\end{equation}
The TFA is obtained by neglecting the
first term in the RHS of Eq.~(\ref{muloc}). Notice that 
the ratio
of the kinetic energy to the interaction energy is proportional to $N^{-2}$.
The fact that in the TFA the local chemical potential scales with 
density as $\pi^2 \hbar^2n^2/2m$, as for a
non-interacting spin-polarized Fermi gas in the local-density
approximation, reflects the property that the hard-core Bose gas has
the same spatial profiles as those of the Fermi gas.

The boson-fermion mapping has also been extended to time-dependent
phenomena \cite{girardeau,girardeau_td}. Therefore, using the mapping {\it in the
inverse direction} we expect that Eq.~(\ref{gp4}) describes
approximately a non-interacting Fermi gas within a local-density
approach. Indeed, it is easily shown that 
upon neglecting the kinetic energy term in $\mu_{loc}(x,t)$ 
Eqs.~(\ref{cont}) and~(\ref{eqv}) yield 
the equation of motion for the density
fluctuations of the Fermi gas in the hydrodynamic regime.
The latter is given by \cite{dynf}
\begin{equation}
m\partial_t^2 n(x,t)=
\partial_x^2 \Pi
(x,t)+\partial_x\left[n(x,t) \partial_x V(x,t)\right]
\label{pinco}
\end{equation}
where the momentum flux density $\Pi(x,t)$ depends locally on the
particle density through the relation
\begin{equation}
\Pi(x,t)=\hbar^2\pi^2 n^3(x,t)/3m+mn(x,t)v^2(x,t)/2\;.
\label{pallino}
\end{equation}
Equations (\ref{pinco}) and (\ref{pallino}) are obtained quite
straightforwardly by combining Eqs. (\ref{cont})-(\ref{muloc}) in the TFA.

Having established that  Eq.~(\ref{gp4}) describes in the TFA limit
the dynamics of both hard-core bosons in the
strong-coupling regime 
and non-interacting spinless fermions in the hydrodynamic regime, we turn
to its equilibrium solution, which is a necessary preliminary step for
obtaining the collective modes.

\begin{figure}
\centerline{\psfig{file=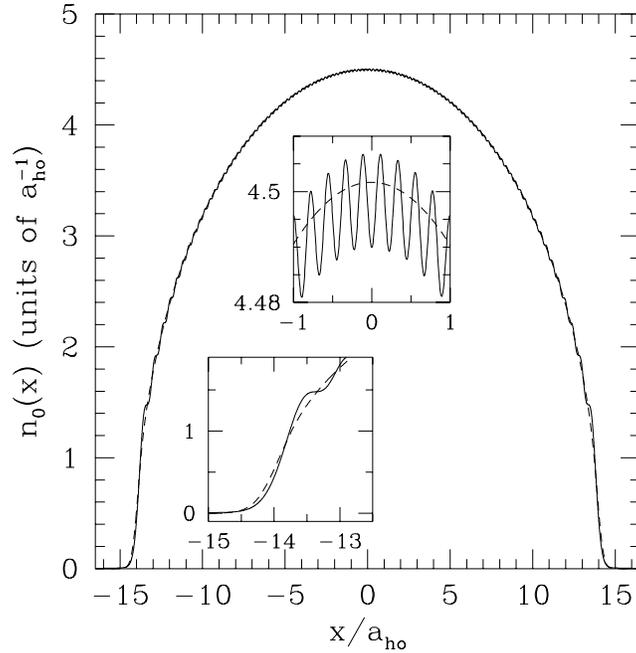,width=0.5\linewidth}}
\caption{Equilibrium density profile for $N=100$ fermions, in units of
the harmonic oscillator length
$a_{ho}=\sqrt{\hbar/m\omega_{ho}}$. The exact profile (solid line) 
is compared with that  obtained from the non-linear
Schr\"odinger equation~(\ref{gp4}) (dashed line). The insets show
enlargements of the region around $x=0$ and of the tail region, in the
same units.}
\label{fig1}
\end{figure}

The ground-state density $n_0(x)$ is obtained from Eq.~(\ref{eqv}) by setting
$\mu_{loc}(x)=\mu$ with $v=0$ and $U_p=0$.
 This yields the following differential equation for the
equilibrium density:
\begin{equation}
-\frac{\hbar^2}{2m}\partial_x^2 \sqrt{n_0(x)} +\frac{\pi^2\hbar^2}{2m}
 n_0^{5/2}(x) +[V_{ext}(x)-\mu]\sqrt{n_0(x)}=0\;.
\label{gp4eq}
\end{equation}
It is important to remark that Eq. (\ref{eqv}) does
not yield the exact equilibrium density profile of the gas 
even if one includes the kinetic energy term 
$(\hbar^2/2m)\partial_x^2 \sqrt{n_0(x)}$. The exact profile satisfies instead
the third-order differential equation \cite{vandoren}
\begin{equation}
 -\frac{\hbar^2}{8m}\partial_x^3 n_0(x)
  -\frac{1}{2}n_0(x)\partial_x V_{ext}(x)
  +[V_{ext}(x)-\mu]\partial_x n_0(x)=0\;.
\label{realeq}
\end{equation}
In  the case of $N=100$ particles an accurate comparison between the
solutions of Eq.~(\ref{gp4eq}) 
(obtained by the numerical method described in Sec.~\ref{sim} below)
and of Eq.~(\ref{realeq}) (using the results of Ref.~\cite{noi})
is given in Fig.~\ref{fig1}. It is seen that although the overall
shape of the two 
curves is quite similar, Eq.~(\ref{gp4eq}) misses the shell structure
of the exact profile and has an incorrect behavior in the
tails (see the insets in Fig.~\ref{fig1}) (see also
Ref.~\cite{kolomeiski}).  

The above comparison illustrates  
the limits of validity of Eq.~(\ref{gp4}). We shall see in the
following that the spill-out of the particle density beyond the
classical radius plays a crucial role in determining the dynamics of
the atomic cloud. We may also remark that, whereas Eq. (\ref{gp4}) is
constructed from an approximation for the kinetic energy density
functional, the exact functional is in fact explicitly known for this
system \cite{vandoren}. However, it involves a 
complicated non-local function of the particle
density.

\section{Collective modes  in the Thomas-Fermi limit}
The linearized equation of motion for small-amplitude density
fluctuations is
\begin{equation}
m\partial_t^2 \delta n(x,t)=\partial_x^2 \delta \Pi(x,t)+\partial_x
[\delta n(x,t) \partial_x V_{ext}(x)]\; ,
\label{fiore}
\end{equation}
where $\delta n(x,t)=n(x,t)-n_0(x)$ and we have assumed resonance
conditions, {\it i.e.} $U_p(x,t)=0$. In the homogeneous gas
$V_{ext}(x)=0$ and Eq. (\ref{fiore}) yields an acoustic dispersion
relation with the propagation velocity $v_F$ of hydrodynamic sound. It
is a peculiarity of the ideal 1D Fermi gas that this coincides with
the velocity of collisionless sound.

In the strict TFA limit the equilibrium density profile has finite 
extension, as it is given by
\begin{equation}
n_0^{TF}(x)=\frac{\sqrt{2m}}{\pi\hbar}\theta(X_F^2-x^2)[\mu-V_{ext}(x)]^{1/2} 
\label{tfaeq}
\end{equation}
where the chemical potential $\mu=\hbar \omega_{ho} N$ is obtained
from normalization of Eq.~(\ref{tfaeq}) to the number $N$ of particles
in the trap. We define the classical turning point as the Fermi radius
$X_F=\sqrt{2N} a_{ho}$ with $a_{ho}=\sqrt{\hbar/m\omega_{ho}}$.  
Correspondingly the TFA expression for the fluctuations in the
momentum flux density is 
$\delta
\Pi_{TFA}(x,t)=2\theta(X_F^2-x^2) [\mu-V_{ext}(x)]\delta n(x,t)$. As
will be apparent from the arguments given below, the spectrum of
collective excitations is obtained as a {\it continuum} if this
expression is used in Eq. (\ref{fiore}).

A {\it discrete} spectrum is instead obtained if the spill-out outside
the classical radius is taken into account. To do this, we adopt the
approximate relation
\begin{equation}
\delta
\Pi(x,t)\simeq 2\frac{\delta t}{\delta n}\delta n(x,t)=
2[\mu-V_{ext}(x)]\delta n(x,t)
\label{tizio}
\end{equation}
where $t(n)$ is the kinetic energy density and the second equality
follows from using the Euler equation for Density Functional Theory.

By substituting Eq. (\ref{tizio}) into Eq. (\ref{fiore}) and
performing a Fourier transform with respect to time, we rewrite Eq. (\ref{fiore})
as the eigenvalue equation
\begin{equation}
(1-z^2)\partial_z^2\delta n(z,\omega)-3z \partial_z
\delta n(z,\omega)+[(\omega/\omega_{ho})^2-1]\delta n(z,\omega)=0\;,
\end{equation}
where $z=x/X_F$ is a rescaled position variable. It can be checked by
direct substitution that for $|z|<1$ two independent solutions of this
second-order differential equation are
\begin{equation}
\delta n_{in}^{(1)}(z,\omega)=\frac{\cos[(\omega/\omega_{ho}){\rm arccos}
z]}{\sqrt{1-z^2}} 
\label{solbin}
\end{equation}
and
\begin{equation}
\delta n_{in}^{(2)}(z,\omega)=\frac{\sin[(\omega/\omega_{ho}){\rm arccos}
z]}{\sqrt{1-z^2}}
\label{solcin}
\end{equation}
for {\it any} value of the ratio $\omega/\omega_{ho}$. In the domain 
$|z|>1$ we instead have 
\begin{equation}
\delta n_{out}(z,\omega)=\frac{\left(|z|-\sqrt{z^2-1}\right)^{\omega/\omega_{ho}}}
{\sqrt{z^2-1}}\; ,
\label{solbout}
\end{equation}
having imposed that the density fluctuations vanish for
$|z|\rightarrow\infty$. Finally, the dispersion relation and the shape
of the density fluctuation modes inside the classical radius are
obtained by imposing continuity of $\sqrt{|1-z^2|}\delta n(z,\omega)$
across the Fermi radius. This condition selects the integer values of
the frequency ratio, 
\begin{equation}
\omega/\omega_{ho}=n
\label{sempronio}
\end{equation}
and the form (\ref{solbin}) of the inner solution. This can also be
written as 
\begin{equation}
\delta n_{in}^{(1)}(z,n\omega_{ho})=\frac{T_n(z)}{\sqrt{1-z^2}}
\label{caio}
\end{equation}
where the functions $T_n(z)$ are the Chebyshev orthogonal polynomials
of the first kind~\cite{grad}.
The (integrable) divergence in Eq. (\ref{caio}) at the Fermi radius is
a consequence of the linearized TFA in low dimensionality and is
evidently inconsistent with the linearization of the hydrodynamic
equations. We shall return on this point in the next Section.

In the case $n=1$ we correctly recover from Eq. (\ref{sempronio}) the
frequency of the sloshing mode, in agreement with the generalized Kohn
theorem \cite{kohn}. More generally, we find that the TFA spectrum
coincides with the spectrum of the Fermi gas in the collisionless
regime, as evaluated in Ref. \cite{VMT}. Although this result for the
trapped gas has been reached by an approximate and to some extent
{\it ad hoc} argument, it will be verified immediately below by direct
numerical solution of Eq. (\ref{gp4}).

\section{Numerical simulation}
\label{sim}
The divergence of the density fluctuations at the classical
boundary of the cloud that result from the extended TFA 
are incompatible with a linearization of the equations of
motion. We have therefore carried out a numerical solution of
Eq.~(\ref{gp4}), which reproduces an experimental
procedure employed to excite collective modes \cite{coll_exp}.
Namely, we first apply a 
time-dependent perturbing field of given symmetry for variable
amounts of time corresponding to several periods of the expected
excitation, and then monitor 
the evolution of the cloud by recording a series of ``pictures'' 
once the perturbation is turned off.
From the 
evolution of the density profiles we obtain the frequency and the
shape of the density fluctuation modes.

The numerical simulations use an explicit time-marching algorithm
\cite{algo} for propagating the state of the cloud in both imaginary and real time, yielding
the ground-state profile $n_0(x)$ and the subsequent dynamical
behavior $n(x,t)$. Due to the size of the
nonlinear term in Eq.~(\ref{gp4}), special attention  has been given to the
stability of the algorithm. The 
perturbing potential has
the form  $U_p(x,t)=U_{p}(x)\cos(\omega_i t)$ and the drive frequency $\omega_i$
is tuned around the expected resonances.
As to the spatial variation of the perturbation, we have
adopted  the TFA form $U_p(x)\propto \sqrt{1-(x/X_F)^2}\delta
n_{in}^{(1)}(x,\omega_i)$  in order to excite well defined modes. This is
approximately equivalent to imposing
orthogonality between the drive field and
the other density fluctuations, in view of the orthogonality of the
Chebyshev polynomials in the domain $|z|<1$. 

\begin{figure}
\centerline{\psfig{file=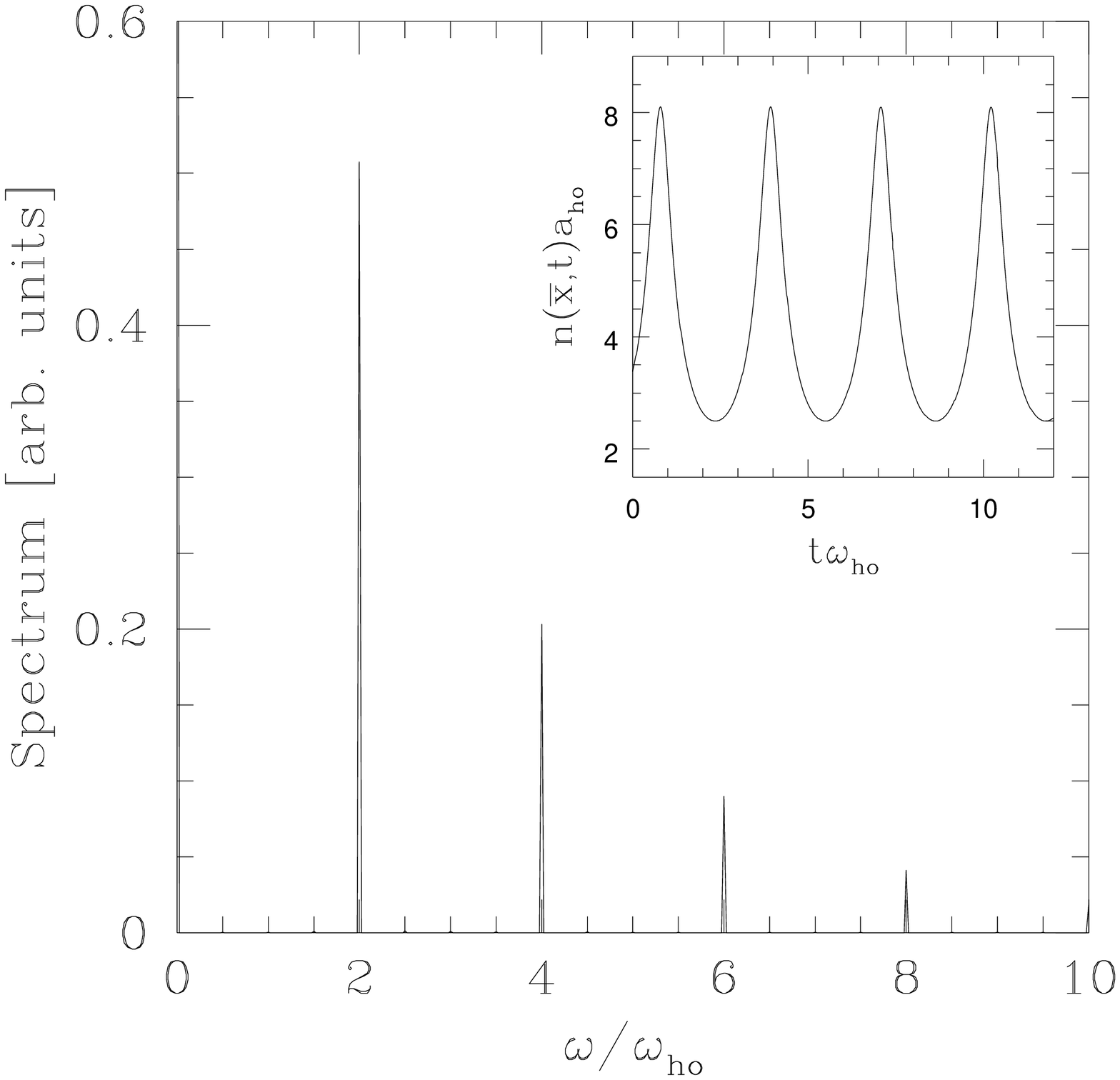,width=0.45\linewidth}\psfig{file=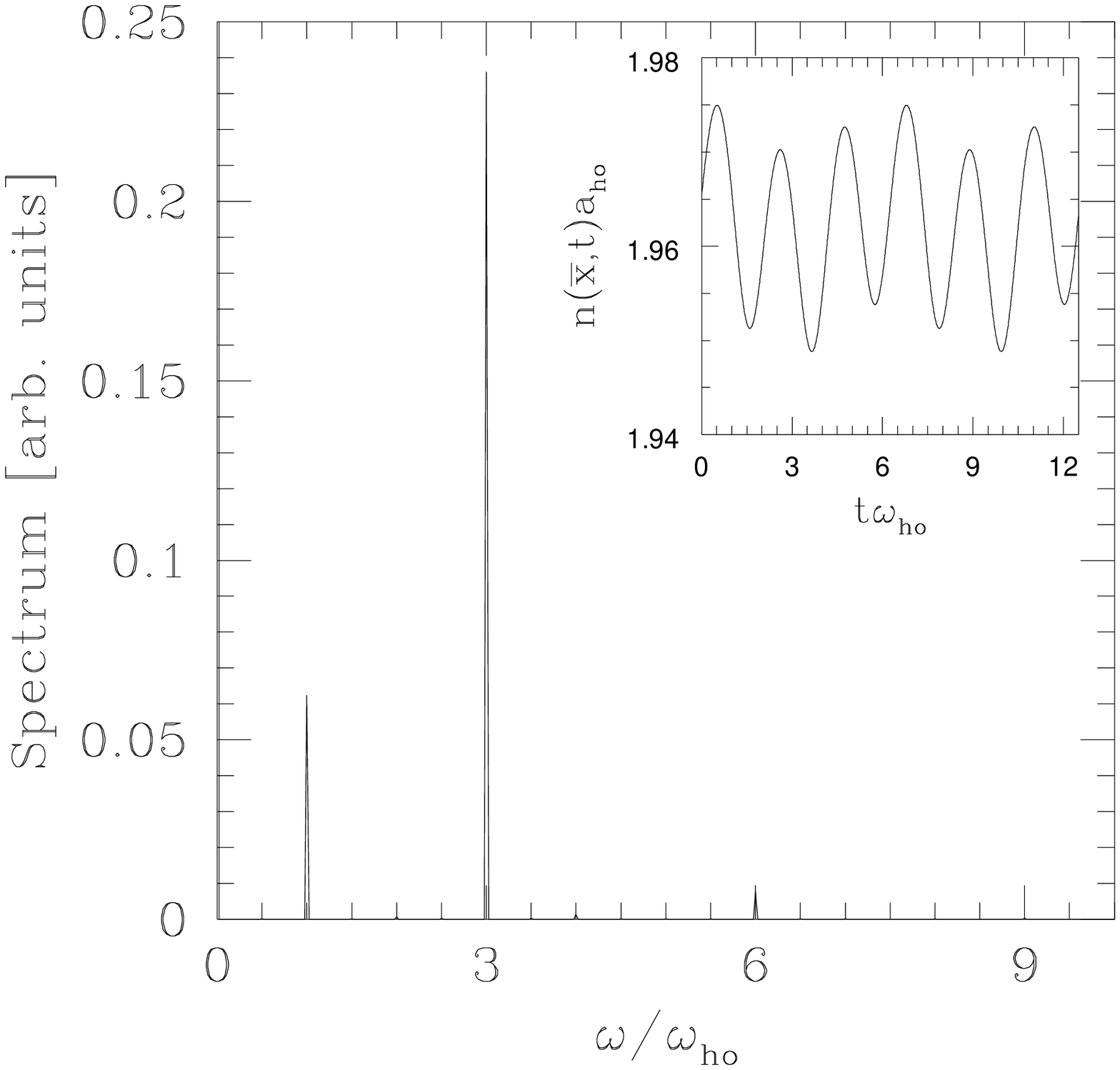,width=0.45\linewidth}} 
\caption{Spectra of density fluctuations 
obtained from the numerical simulation by applying a
perturbing potential of drive frequency $\omega=2$ $\omega_{ho}$ (left
panel) and $\omega=3$ $\omega_{ho}$ (right panel) and of the appropriate
symmetry as described in the text. The insets show the corresponding 
time evolution of the density profile $n(\bar x,t)$ (in units of
$a_{ho}^{-1}$, with $t$ in units of $\omega_{ho}^{-1}$) 
taken at a given point $\bar x$ in the
trap. The spectra are obtained from the
Fourier transform of $n(\bar x,t)$, extended
over several periods for a better definition of the mode
frequencies.} 
\label{fig2}
\end{figure}

Figure~\ref{fig2} shows the spectra of quadrupole and hexapole 
collective modes for $N=100$ particles 
 as obtained from the Fourier transform of the time
evolution of the density profile taken at a given point in the harmonic
trap. The results 
agree with the TFA prediction that the main frequencies of these
modes are $\omega=2$ $\omega_{ho}$ and $\omega=3$ $\omega_{ho}$,
respectively. Nonlinearity is apparent through the presence of several
other resonances in each spectrum. 
We have checked that the spectra of the $i^{th}$ moments
$\langle x^i\rangle=(1/N)\int dx\, x^i n(x,t)$ are
peaked at the fundamental excitation frequency of each mode. 
In further tests we have ({\it i}) examined the behavior of the quadrupole mode
frequency with varying $N$, finding only very small deviations within
the error bars from the value $\omega=2\omega_{ho}$ at low $N$; and
({\it ii}) used non-integer 
values for the driving frequency, finding again only spectral lines at
integer frequencies.

\begin{figure}
\centerline{\psfig{file=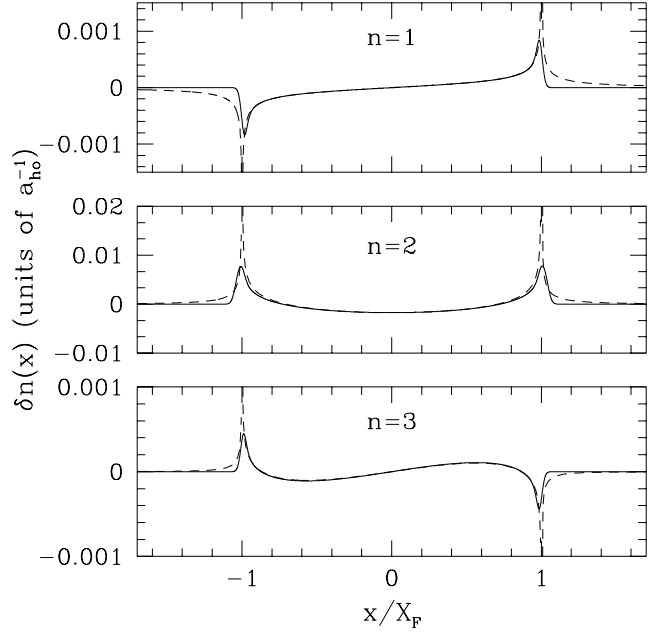,width=0.55\linewidth}}
\caption{Density fluctuation profiles $\delta n(x)$ 
for the 
low-lying modes (indicated by the mode numbers $n=1$, 2 and 3 from top
to bottom) at 
$N=100$ particles, in units of
$a_{ho}^{-1}$ as functions of position $x$ in units of the
Thomas-Fermi radius $X_F$. The results from the numerical simulation (solid lines),
where $\delta n(x)=n(x,\bar t)-n_0(x)$ and $\bar t$ is a given time,
are compared with the 
analytic solutions in the Thomas-Fermi limit (dashed lines).   
}
\label{fig3}
\end{figure}

Finally, in a regime of weak drive amplitude we have compared
the density fluctuation profiles in Eqs.~(\ref{solbin})
and~(\ref{solbout}) with the output of the
simulation. The comparison is illustrated in
Fig.~\ref{fig3}. The TFA density fluctuations, in addition 
to suffering from divergiencies at the classical radius, do not
integrate to zero for the modes with $n=$even integer. 

\section{Summary and conclusions}
In summary, we have shown by analytical and numerical arguments that
the hydrodynamic frequency spectrum of the 1D ideal Fermi gas under
harmonic confinement is given by integer multiples of the trap
frequency. This result extends to the trapped Fermi gas a property of
the homogeneous ideal Fermi gas, in which the speed of sound is the
same in the hydrodynamic regime as in the collisionless regime.
A more accurate treatment of the collisional
regime beyond the TFA may usefully be
developed in the future.

A further conclusion can be drawn from the present solution of the
non-linear Schr\"odinger equation describing a 1D
hard-core Bose gas. By virtue of the fermion-boson mapping theorem,
the frequency spectrum associated with this local-density  theory
reproduces the exactly known 
spectrum of the Bose gas under harmonic confinement.

\newpage

\end{document}